\documentclass[]{IEEEtran}
\usepackage{microtype}
\usepackage{graphicx}
\usepackage{xcolor}
\usepackage{stfloats}
\usepackage{subfigure}
\usepackage{fixltx2e}
\usepackage{balance}
\usepackage[square,sort&compress,comma,numbers]{natbib}

\title{LASeR: Lightweight Authentication and Secured Routing for NDN IoT in Smart Cities}
\author{
    \IEEEauthorblockN{Travis Mick, Reza Tourani, Satyajayant Misra}
    \IEEEauthorblockA{\\Dept. of Computer Science, New Mexico State University
    \\\{tmick, rtourani, misra\}@cs.nmsu.edu}
    ~\thanks{This work is supported in part by the U.S. NSF grants:1345232 and 1248109 and the U.S. DoD/ARO grant: W911NF-07-2-0027. 
    {\em Paper currently in submission in IEEE IoT Journal}.}
}
\date{November 2016}

\begin{document}

\maketitle

\begin{abstract}
Recent literature suggests that the Internet of Things (IoT) scales much
better in an Information-Centric Networking (ICN) model instead of the current
host-centric Internet Protocol (IP) model.
In particular, the Named Data Networking (NDN) project (one of the ICN
architecture flavors) offers features exploitable by IoT applications, such
as stateful forwarding, in-network caching, and built-in assurance of data
provenance.
Though NDN-based IoT frameworks have been proposed, none have adequately
and holistically addressed concerns related to secure onboarding and
routing.
Additionally, emerging IoT applications such as smart cities require high
scalability and thus pose new challenges to NDN routing.
Therefore, in this work, we propose and evaluate a novel, scalable framework for
lightweight authentication and hierarchical routing in the NDN IoT (NDNoT).
Our ns-3 based simulation analyses demonstrate that our framework is scalable and efficient.
It supports deployment densities as high as 40,000 nodes/km$^2$ with an average
onboarding convergence time of around 250 seconds and overhead of less than 20 KiB per node. 
This demonstrates its efficacy for emerging large-scale IoT applications such as smart cities.
%
%


\end{abstract}

\section{Introduction}

The new emerging concept of smart cities applies concepts from the Internet
of Things (IoT) to the management of diverse municipal infrastructure and
assets~\citep{zanella2014internet}.
%
%
Smart cities will involve large numbers of IoT devices installed in a range of
settings from individual homes to critical infrastructure, potentially in
a very dense deployment.
As a result, their feasibility will require advances in efficiency and
scalability of IoT communications.
%
%
Additionally, smart cities will require strong guarantees of security: networked
devices will handle large volumes of sensitive information and control valuable
assets such as utility infrastructure, thus widening the attack surface for potential
compromise.
Thus, strong end-to-end security and privacy mechanisms between smart devices and
the cloud are imperative.
Recent literature suggests that Information-Centric Networking (ICN)
is a more appropriate approach than Internet Protocol (IP) for
IoT~\citep{baccelli2014information}.
Named Data Networking (NDN)~\citep{zhang2010named}, in particular, is a strong architecture
for creating scalable and efficient smart city networks, by employing
features such as stateful forwarding and in-network caching.
In addition, it offers security benefits such as enforced provenance through mandatory
network-layer signatures.
Several ICN-based IoT deployments have been announced in the literature, however
no holistic NDN of Things (NDNoT) architecture and protocol suite has yet
been proposed.
In particular, existing literature tends to neglect concerns related to secure routing and
onboarding, two of the most difficult problems in IoT.
Works that do address routing or onboarding do so separately, neglecting the fact that
they are closely coupled.
As a result, the proposed disjoint schemes are either incompatible or inefficient
when combined.
Therefore, we propose in this paper a scalable and secure framework addressing both
onboarding and routing in the NDNoT.
The scalability of an IoT deployment is adversely affected by the low computational and
memory capacities of IoT devices, as well as the characteristics of the low power
lossy networks (LLNs) they use to communicate.
Thus, we employ a hierarchical network structure, a design which has been 
recommended to achieve scalability in IoT~\citep{hong2013mobile}.
Such an architecture allows us to offload the burden of routing onto a few
less-constrained ``anchor'' nodes, while other devices need only form 
destination-oriented trees.
In our framework, secure onboarding is made a prerequisite to routing, in order
to help protect the network against routing attacks such as
blackholes~\citep{kannhavong2007survey}.
Each node in the network is authenticated prior to commencing routing, and
in turn a node also authenticates the network it is joining.
Since asymmetric cryptography is typically infeasible on IoT devices, we use
symmetric cryptography.
%
%
Our onboarding protocol is based on pre-shared keys between each node and a 
designated authentication manager in the infrastructure.
We have combined our approaches to routing and onboarding into a single holistic
framework for Lightweight Authentication \& Secured Routing (LASeR).
The combined authentication and onboarding processes are very lightweight, requiring
only three round trips and few cryptographic operations.
In summary, the \textbf{contributions} of our work are:
(1) We analyze the current state-of-the-art of routing and authentication in the
NDNoT;
(2) We propose LASeR, a holistic framework for efficient and secure onboarding and routing
in NDN; and
(3) We demonstrate LASeR's effectiveness and efficiency through analyses conducted in
ndnSIM, the NDN module for ns-3.
The remainder of this paper is organized as follows: Section~\ref{sec:related} reviews
prior work on NDN and IoT; Section~\ref{sec:system} presents our model for the IoT
network and reviews the primitives employed by NDN; Section~\ref{sec:crypto} describes
the cryptographic materials and operations underlying LASeR's authentication mechanism; 
Section~\ref{sec:protocol} presents the protocols employed for onboarding and routing;
Section~\ref{sec:results} offers a simulation-based validation of LASeR's effectiveness;
and finally, Section~\ref{sec:conclusion} concludes the paper and gives an overview
of our planned future work.

\section{Related Work}
\label{sec:related}

Benefits and challenges related to the ICN-based IoT have been previously discussed
in the literature~\citep{amadeo2014named,datta2016integrating,baccelli2014information}, and
several architectures have been proposed for both general IoT~\citep{baccelli2014information}
and specific applications~\citep{ravindran2013information, amadeo2015information,
burke2012authenticated}.
However, the majority of these designs focus on service discovery, data delivery, and
similar application-centric concerns rather than the initial network bootstrapping or
route discovery procedures and their security.
Though most of the aforementioned works do not suggest novel routing protocols for IoT,
\citep{baccelli2014information} recognized the routing-related challenges imposed by device
constraints in the IoT and proposed a new opportunistic-reactive routing protocol.
Under this model, forwarding tables are populated after observing the origins of downstream
packets; a flooding-based approach is used as a fallback when no proper route is available.
A similar approach was previously outlined in \citep{meisel2010ad}.
Other approaches to ad-hoc routing in NDN were reviewed in \citep{amadeo2015forwarding};
the authors identified two broad classes of routing protocols: provider-blind
and provider-aware.
The provider-blind schemes solely employ controlled flooding to forward requests,
while provider-aware schemes add a reactive mechanism like that in the two designs mentioned
above.
All these routing protocols employ reactive designs; however, we choose a proactive approach
like that of the IPv6 Routing Protocol for Low-Power and Lossy Networks (RPL)~\citep{rfc6550},
which is currently favored for the IP-based IoT.
Bootstrapping and onboarding for the ICN-based IoT have only recently been given
serious consideration.
Previous architectures such as \citep{burke2012authenticated} relied on the asymmetric
authentication mechanisms used throughout the NDN stack, however \citep{enguehard2016cost}
quantified the time and energy overheads of such schemes on constrained devices and
ultimately concluded that their cost is too high.
As a result, two designs based on symmetric cryptography were proposed in
\citep{compagno2016onboardicng}: a basic implementation of the authenticated key
exchange protocol (AKEP2)~\citep{bellare1993entity} over ICN, and an improved version
which increases its efficiency.
Though \citep{compagno2016onboardicng} efficiently addresses the initial
authentication and key-distribution challenges for IoT, it does so without
regard to the needs of a routing protocol.
As a result, to employ it in conjunction with a separate routing framework would
impose additional overhead; the two steps of authentication and routing will occur serially, 
increasing overhead in the network and overall onboarding latency.
In light of this, we propose LASeR, wherein elements of authentication and routing can occur
simultaneously to reduce their overall cost.

\section{System and Threat Models and Assumptions}
\label{sec:system}
In this section, we present the system, network, and threat models and assumptions. 
For better understanding of our models and assumptions, we start with an overview of NDN. 
\subsection{Overview of NDN}
The ``thin waist'' of the Named Data Networking (NDN) stack, as the name implies,
is Named Data.
In the NDN model, each chunk of data (typically referred to as a
\textit{content object}) has a unique \textit{Name}, similar to a Uniform
Resource Identifier (URI); the content associated with each Name is typically
considered to be immutable.
To retrieve a particular content object, a requester sends an \textit{Interest}
packet into the network.
At a minimum, the Interest contains the Name of the desired content object; it can
also contain a signature to verify the requester's identity.
The network then retrieves the appropriate content object and delivers it to the
requester as a \textit{Data} packet.
The Data contains, at a minimum, its Name, the actual content payload, and its
publisher's signature.
The requester can then verify the signature to ascertain the content object's
authenticity.
Both Interest and Data signatures typically (but optionally) include a
\textit{KeyLocator} field, which contains the Name of the key used for
the signature.
Each router in NDN maintains three data structures: a Pending Interest Table (PIT), a
Forwarding Information Base (FIB), and a Content Store (CS).
The forwarding procedures for both Interests and Data are based around these tables.
Upon receiving an Interest, a router first checks its CS for a match;
the CS essentially serves as a cache of Data, indexed by Name.
If a match is found in the CS, the Data is served and the request is considered
satisfied.
If no match is found in the CS, the router then checks its PIT, which indicates
whether a previous Interest for the same Name has been forwarded but not yet
satisfied.

If a PIT entry exists, the router need not forward the Interest again; instead,
it adds the identifier of the incoming interface (\textit{Face}, in the NDN
nomenclature) to that PIT entry.
If no PIT entry is found, the router consults its FIB (essentially a forwarding
table) and employs a configurable \textit{forwarding strategy} to identify the
correct Face on which to forward the Interest.
The router then adds a new PIT entry indicate that the Interest was forwarded.
Data packets are essentially forwarded following the reverse path as indicated
by matching PIT entries.
That is, a router receiving a Data checks its PIT to determine the correct Face(s)
on which to forward the Data.
Once the Data is forwarded, the PIT entry is cleared.
The Data may then be added to the CS and used to satisfy future requests, depending
on the policy employed for cache admission and eviction.
Note that the configurability of the forwarding strategy is an important
feature for the application of NDN in IoT.
A different strategy can be employed for each Interest depending on its Name prefix,
allowing, for example, enhanced Quality of Service (QoS) depending on the nature of
the request.
In LASeR, we employ a custom strategy to facilitate our hierarchical network
design; more details on this strategy are given in Section~\ref{subsec:forwarding}.

\subsection{System Model and Assumptions}

\begin{figure}[]
	\centering
	\includegraphics[height=2.4in]{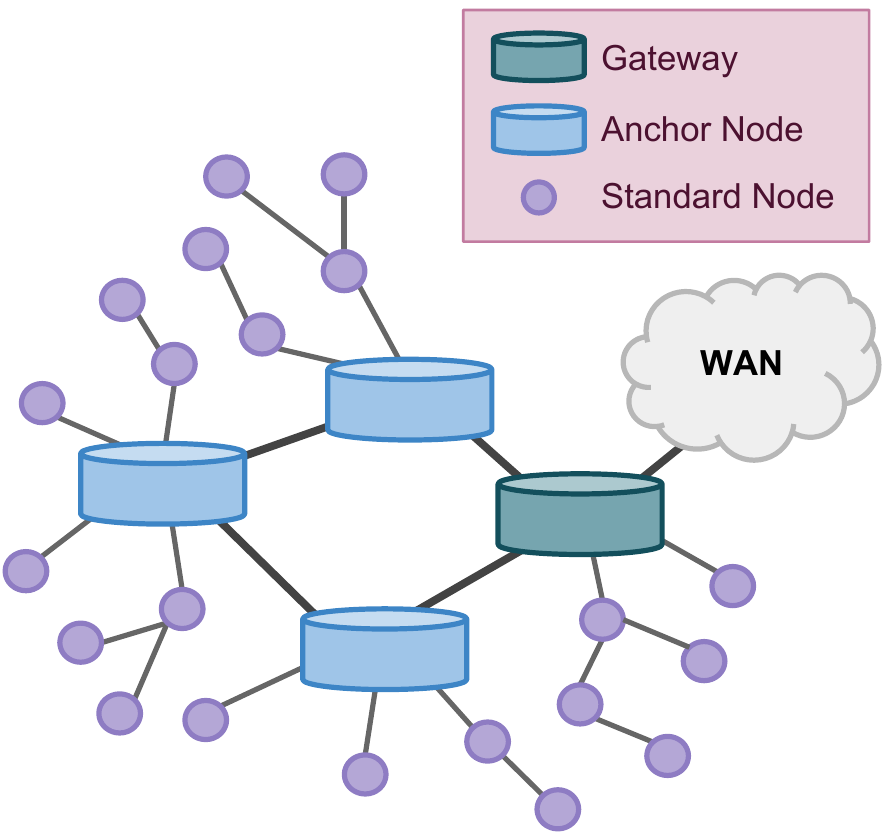}
	\caption{
		\label{fig:system}
		In our hierarchical island, a gateway connects to the WAN,
		anchors form an island backbone, and standard nodes form trees
		rooted at the anchors.
	}
\end{figure}

We model the NDNoT as consisting of \textit{islands}, which exist at the edge
of the greater Internet.
The protocols employed within the island need not be influenced by those used
in the wide-area network (WAN); therefore, this model is suitable for a
local clean-slate deployment of NDN in smart cities prior to wide
adoption.
We distinguish between three types of nodes within each island:
gateways, anchor nodes (ANs), and standard nodes (SNs).
We assume that SNs have small memory, computation, and energy capacities, and
employ LLN radios; on the other hand, gateways and ANs are essentially unconstrained.
The connections between these entities are visualized in Fig.~\ref{fig:system}.
Gateways serve as edge routers between the island and the WAN, and 
the ANs are a superset of the gateways and form a backbone or core
for the island.
Standard nodes wirelessly peer with ANs and use them as sinks to facilitate
communication, thus creating trees, or \textit{clusters}, of constituent SNs
around each AN.
We assume each SN is assigned a flat identifier (ID), which could either be derived
from its media access control (MAC) address or be chosen arbitrarily.
For scalability, we will use these IDs to perform routing and forwarding.
Nodes can also advertise arbitrary, application-specific Name prefixes; other
requesters would then resolve these Names into IDs for the purpose of routing.
Namespace creation and management is an NDN- and application-specific decision.
This is outside the scope of this work.

In addition to network entities named above, we assume that there is a service
capable of managing the authentication and registration of nodes in the network.
We will refer to this entity as the Island Manager (IM); it may exist either in the
cloud, within some particular node, or even as a synchronized database shared between
anchors.
We assume that the IM and ANs are synchronized to perform secure communication and routing. We do not discuss mechanisms to achieve this, however it is easy to design and implement.
The placement of the IM is an implementation detail which should be made
with consideration to the specific needs of a particular deployment.
The IM will be responsible for node authentication, and will also serve Name-to-ID
resolution requests to support hierarchical routing.

\subsection{Threat Model and Assumptions}

We assume that all the devices in the network are capable of performing symmetric key cryptography, 
such as advanced encryption standard (AES), and message authentication using keyed-hashed functions,
such as hashed-MAC. 
As is standard, we assume that the encryption algorithms and the MAC functions cannot be compromised.    
In our system, there can be both inside and outside attackers. 
An outside attacker is not part of the network.
It can passively capture data transmissions in the network to perform traffic analysis and also replay
captured packets. 
It can also be an active attacker attempting to masquerade as a legitimate node, and can try to inject false
data into the network. 
An inside attacker is a node that is already on-boarded into the network, it can also inject false data
in the network. 
The false data can include fake route advertisements, enabling sinkhole or blackhole attacks.
A compromised or colluding node's keying materials can be extracted and used by an adversary, not part of
the system, to impersonate  as a legitimate node.
This is termed sybil attack; the compromised adversary can operate as a legitimate node in the network. 
Denial of service and channel jamming can also be threats in our system. 

\section{Cryptographic Materials and Primitives}
\label{sec:crypto}

\subsection{Overview}

\begin{figure}[]
\centering
\includegraphics[height=2.4in]{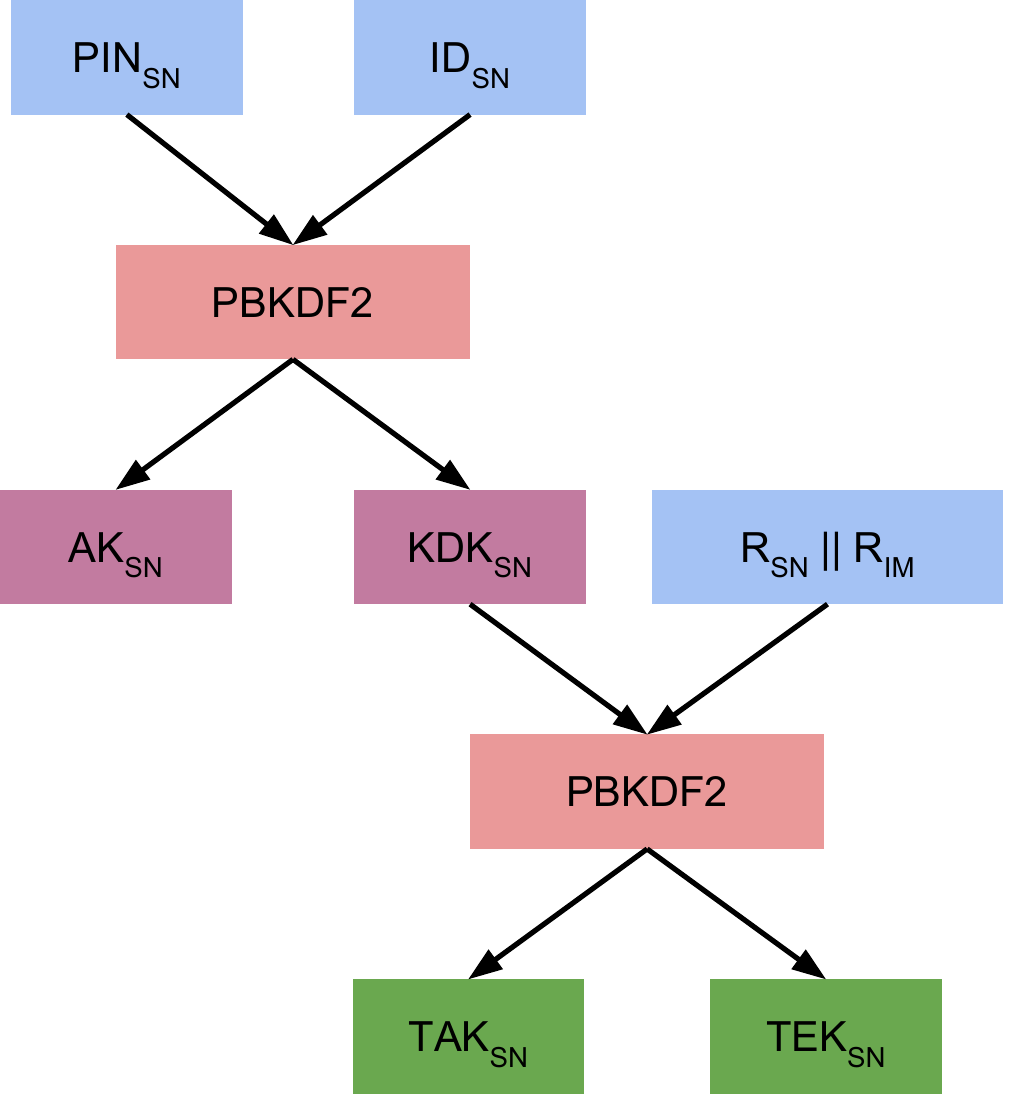}
\caption{
	\label{fig:keys}
	LASeR's key derivations are based around a pre-shared key,
	PIN\textsubscript{SN}.
	Two permanent keys are shared, and two transient keys are derived per session.
}
\end{figure}

The key hierarchy of LASeR, visualized in Fig.~\ref{fig:keys}, is inspired by that
of the Pre-Shared Key Extensible Authentication Protocol (EAP-PSK) \citep{rfc4764}.
A session between an SN and an IM is identified by the respective IDs of the two
parties as well as two nonces (one chosen by each).
The SN and IM initially share a pre-shared key (PSK), from which two long-lived
keys are derived (one for key derivation, one for authentication).
With the exchange of nonces and establishment of a session, two additional
transient keys are established (one for encryption, one for authentication).
These transient keys can be intermittently refreshed simply by exchanging new nonces.
%
%
%
%
%

\subsection{Permanent Materials}

Each SN is required to store at least two permanent pieces of information: its
ID (ID\textsubscript{SN}) and its PSK (PIN\textsubscript{SN}), which could be
installed at the time of manufacture.
The IM is required to permanently store only its own ID (ID\textsubscript{IM}).
The IDs may be arbitrary, and the PSK should be random.

\begin{figure}[]
	\includegraphics[width=\columnwidth]{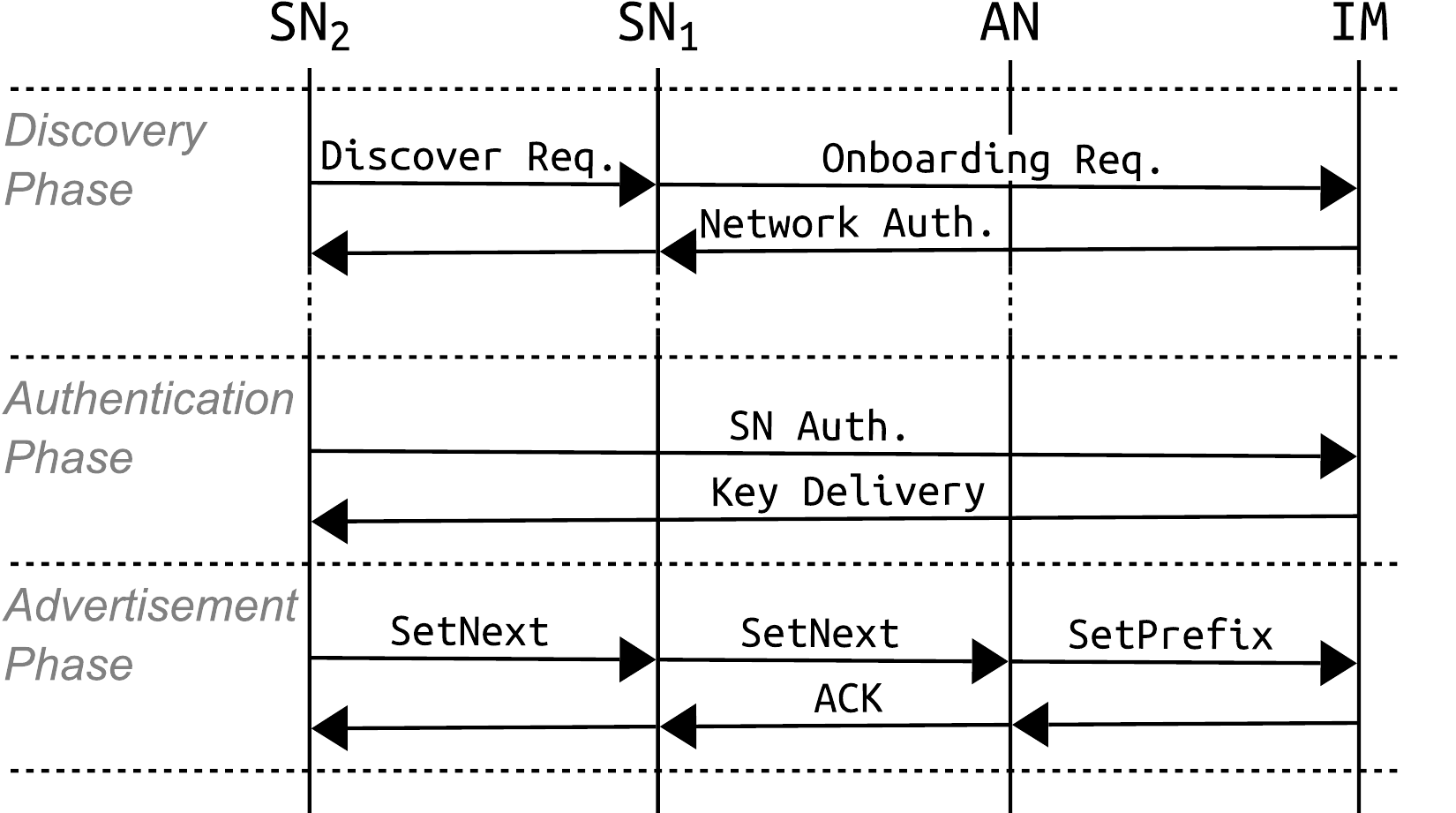}
	\caption{
		\label{fig:proto_overview}
		LASeR consists of three phases, each involving one round-trip
		to the IM.
		The discovery phase can be iterated to obtain a desirable path.
	}
\end{figure}

\subsection{Long-Lived Keys}
\label{subsec:long_keys}

As in EAP-PSK, we use a PSK (in this case, PIN\textsubscript{SN})
in order to derive two long-lived keys: the Authentication Key (AK\textsubscript{SN}),
and the Key-Derivation Key (KDK\textsubscript{SN}).
For ease of implementation, we use a password-based key derivation function
(PBKDF2)~\citep{rfc2898}, rather than the modified counter mode block cipher used by 
EAP-PSK, to derive these keys from the PSK.
Following the construction in EAP-PSK, we configure PBKDF2 with the following
options: PIN\textsubscript{SN} as the password, ID\textsubscript{SN} as the salt,
and an output length of 256 bits.
The first 128 bits of output shall be used as AK\textsubscript{SN}, and the last 128
bits as KDK\textsubscript{SN}.
We use HMAC-SHA256 as the pseudorandom function behind PBKDF2, due to its wide use and
ease of implementation.
The SN may optionally pre-generate and cache both AK\textsubscript{SN} and
KDK\textsubscript{SN} permanently, though it is not required to.
The IM cannot generate these keys until binding time, as it may not have
prior knowledge of ID\textsubscript{SN}.
To enable use of NDN's in-stack authentication features, AK\textsubscript{SN} should
be registered on both nodes as \texttt{/keys/<ID\textsubscript{SN}>/AK}, and
KDK\textsubscript{SN} registered as \texttt{/keys/<ID\textsubscript{SN}>/KDK}.
%
%

\begin{figure*}[t!]
	\centering
	\includegraphics[width=.85\textwidth]{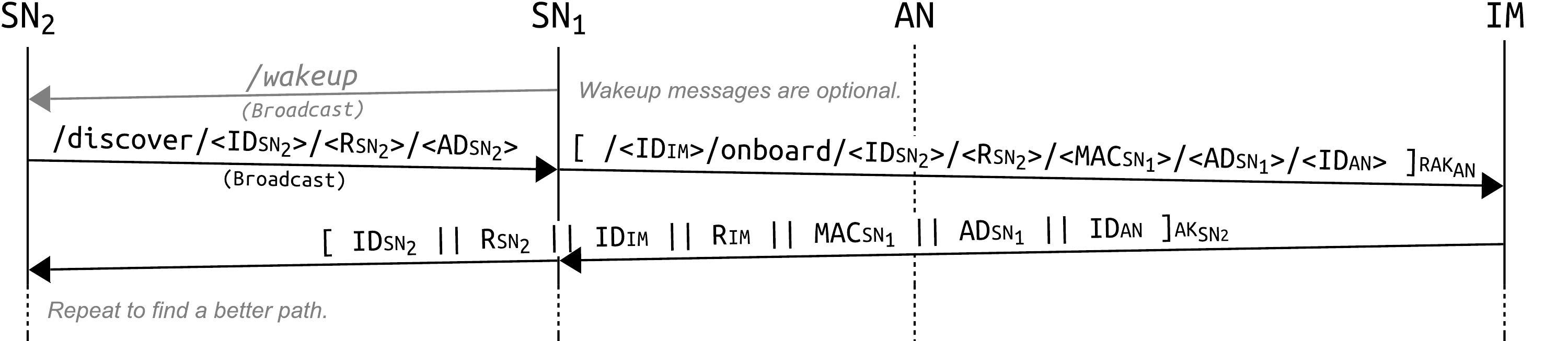}
	\caption{
		\label{fig:proto_discovery}
		The first phase of LASeR involves SN\textsubscript{2} discovering a network
		through its neighbor SN\textsubscript{1}.
		The IM authenticates the network to SN\textsubscript{2} using its knowledge of 
		PIN\textsubscript{SN\textsubscript{2}}.
		The AN (indicated by dotted line) is not involved in the protocol, however
		messages between SN\textsubscript{1} and IM will pass through it en route.
	}
\end{figure*}

\subsection{Transient Keys}

To enhance security, the static keys derived directly from the PSK are not
used to transmit application data, but only to bootstrap the authentication
process.
As in EAP-PSK, two nonce-based ephemeral keys will be derived from the
key-derivation key.
In particular, KDK\textsubscript{SN} is used to derive two transient keys: a
Transient Authentication Key (TAK\textsubscript{SN}), and a Transient Encryption
Key (TEK\textsubscript{SN}).
Again, we use PBKDF2 with HMAC-SHA256 to derive 256 bits of keying material.
The key KDK\textsubscript{SN} is used as the password, and a pair of nonces
(R\textsubscript{SN} and R\textsubscript{IM}) established during the handshake
is used as the salt (details in Section~\ref{sec:protocol}).
These keys are also registered within the local NFD for ease of use:
TAK\textsubscript{SN} is registered as
\texttt{/keys/<ID\textsubscript{SN}>/<R\textsubscript{SN}>/<R\textsubscript{IM}>/TAK},
and TEK\textsubscript{SN} as
\texttt{/keys/<ID\textsubscript{SN}>/<R\textsubscript{SN}>/<R\textsubscript{IM}>/TEK}.

\begin{figure}[b]
	\centering
	\includegraphics[width=.95\columnwidth,trim={0.5cm 0 0.5cm 0}, clip]{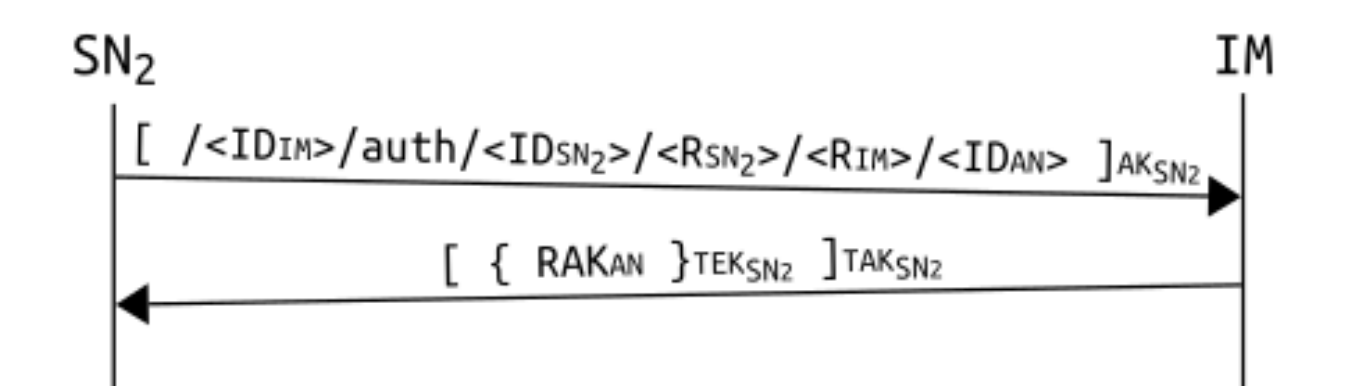}
	\caption{
		\label{fig:proto_authentication}
		In the second stage of LASeR, SN\textsubscript{2} authenticates itself
		to the IM and obtains the RAK corresponding to its anchor.
	}
\end{figure}

\subsection{Secure Channel}

Once TAK\textsubscript{SN} and TEK\textsubscript{SN} are derived, a secure channel
can be established.
Messages are encrypted with AES128-CBC under TEK\textsubscript{SN}, while 
TAK\textsubscript{SN} is used for HMAC-SHA256 signing.
These keys are used to deliver an additional symmetric key to used
for route advertisement, namely the Routing Authentication Key (RAK).
In the following sections, we use the notation \texttt{[~...~]\textsubscript{K}}
to indicate that a message is signed under the key \texttt{K}, and 
\texttt{\{~...~\}\textsubscript{K}} to indicate that a message is encrypted under
\texttt{K}.

\section{The LASeR Protocol}
\label{sec:protocol}

\subsection{Overview}

Onboarding and routing using LASeR occurs in three steps, depicted in
Fig.~\ref{fig:proto_overview}: (1) network discovery and authentication,
(2) SN authentication and key delivery, and (3) path advertisement.
In the first phase, an SN discovers an already-onboarded neighbor, who then
asks the IM for the information necessary to authenticate the network to the new SN.
In the second phase, the SN authenticates itself to the IM and acquires the keys
necessary to advertise a route.
The final phase consists solely of the SN advertising its route; the route is then
propagated hop-by-hop toward the anchor using SetNext messages. The anchor then notifies the IM of the SN's registration using a SetPrefix message.
The full process can be performed in as little as three round trips between
a joining SN and the IM.
The resulting routes from SNs to ANs are similar to those which would be obtained
by a scheme based on destination-oriented directed acyclic graphs (DODAGs) such as
RPL~\cite{rfc6550}.
However, each node chooses only one upstream path in LASeR and therefore the
result is a forest of trees, each rooted at an AN (equivalent to sinks in RPL
nomenclature).
Routing between anchors and gateways is assumed to be handled by other means,
e.g. a link-state protocol, and is beyond the scope of this work.

\subsection{Network Discovery and Authentication}

The first stage of LASeR is network discovery and authentication; that is, an SN discovers
a path and verifies the legitimacy of the network it is connecting to.
It involves three network entities: the node joining the island (SN\textsubscript{2}),
its neighbor (SN\textsubscript{1}), and the island manager (IM).
The discovery process may begin either when SN\textsubscript{2} first comes online,
or at a later time when it receives a wakeup beacon (a notification from a
newly-onboarded neighbor that a path to an AN is now available).
The complete discovery protocol is presented in Fig.~\ref{fig:proto_discovery}.
The first transmission in this phase is a \textit{Discovery Request} sent by SN\textsubscript{2},
which constitutes a request to join an island.
This transmission is an Interest under the \texttt{/discover/} prefix, which is assumed to be
broadcast-forwarded in order for SN\textsubscript{2} to identify an immediate neighbor.
The Interest should have a relatively long PIT lifetime (likely on the order of minutes), as it may
require human input at the IM (to enter PIN\textsubscript{SN\textsubscript{2}}, if it is not
pre-shared) before a Data can be sent in response.
This initial Interest sent by SN\textsubscript{2} contains its ID (ID\textsubscript{SN\textsubscript{2}}), a self-generated nonce (R\textsubscript{SN\textsubscript{2}}), and its current hop-distance
from an anchor (AD\textsubscript{SN\textsubscript{2}}, initially $\infty$); the complete name is
\texttt{/discover/<ID\textsubscript{SN\textsubscript{2}}>/<R\textsubscript{SN\textsubscript{2}}>/<AD\textsubscript{SN\textsubscript{2}}>}.
Any neighbor (SN\textsubscript{1}) which receives this message, is fewer than AD\textsubscript{SN\textsubscript{2}}$ - 1$ hops from an anchor, and wishes to serve as a relay
for SN\textsubscript{2} shall relay it to its AN along with its own MAC
(MAC\textsubscript{SN\textsubscript{1}}), its hop-count distance from an anchor
(AD\textsubscript{SN\textsubscript{1}}), and the ID of that anchor (ID\textsubscript{AN}).
This message, an \textit{Onboarding Request}, essentially represents SN\textsubscript{1}'s assent to
providing a route towards AN for SN\textsubscript{2}.
To this end, SN\textsubscript{1} constructs a new Interest for
\texttt{/<ID\textsubscript{IM}>/onboard} \texttt{/<ID\textsubscript{SN\textsubscript{2}}>/<R\textsubscript{SN\textsubscript{2}}>/<MAC\textsubscript{SN\textsubscript{1}}>/<AD\textsubscript{SN\textsubscript{1}}>/<ID\textsubscript{AN}>} and signs it under RAK\textsubscript{AN} (which is shared by all successfully-onboarded nodes under the AN, as well as by the IM).

\begin{figure*}[t!]
	\centering
	\includegraphics[width=.98\textwidth,trim={1em 0 2em 0}]{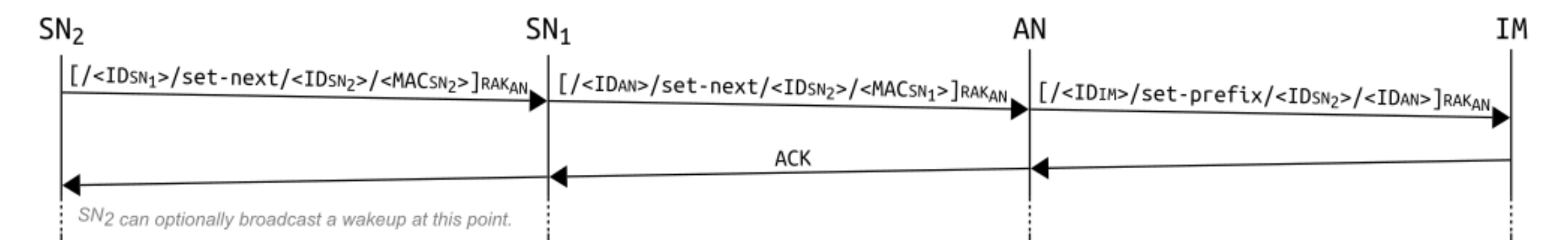}
	\caption{
		\label{fig:proto_advertisement}
		In the final stage of LASeR, SN\textsubscript{2} notifies its neighbor SN\textsubscript{1}
		of its commitment to its path.
		SN\textsubscript{1} then sends a similar notification to the next hop; this repeats
		until AN learns a route to SN\textsubscript{2}.
		Then, AN informs IM that it is serving as the anchor for SN\textsubscript{2}.
		Note that the IM may be co-located with the AN.
	}
\end{figure*}

Upon receiving this Interest, the IM derives AK\textsubscript{SN\textsubscript{2}}
and KDK\textsubscript{SN\textsubscript{2}} according to the procedure outlined in Section~\ref{subsec:long_keys}.
It generates its nonce R\textsubscript{IM} and replies with a \textit{Network Authentication} (NA)
message, which is a Data containing ID\textsubscript{SN\textsubscript{2}},
R\textsubscript{SN\textsubscript{2}}, ID\textsubscript{IM}, R\textsubscript{IM},
MAC\textsubscript{SN\textsubscript{1}}, AD\textsubscript{SN\textsubscript{1}}, and
ID\textsubscript{AN}.
The Data is signed under AK\textsubscript{SN\textsubscript{2}}.
This Data authenticates IM to SN\textsubscript{2}, informs it of its next-hop
neighbor (SN\textsubscript{1}), its distance from an anchor
(AD\textsubscript{SN\textsubscript{1}} + 1), and its anchor (AN).
Because SN\textsubscript{1} changed the Interest name in-flight, it must perform
the corresponding reverse mapping in order to deliver the message to SN\textsubscript{2};
i.e., the application layer changes the Data's Name from that in the Onboarding Request
to that in the original Discovery Request.
After obtaining this Data, SN\textsubscript{2} may send a new discover Interest in
order to attempt to locate a shorter path to an anchor (in the context of our example,
a different node would then take on the role of SN\textsubscript{1}).
To do so, it sends the same Interest as previously but with a new nonce and an updated
AD field.
This process may be iterated as many times as desired, or until SN\textsubscript{2}
no longer receives a useful response.
When SN\textsubscript{2} is content with its path, it notes its next hop toward the anchor
as ID\textsubscript{SN\textsubscript{1}} and its anchor as ID\textsubscript{AN}, then proceeds
to phase two as follows.

\subsection{SN Authentication and Key Delivery}

After completing the first phase, SN\textsubscript{2} trusts its island (via its trust for
the IM) and is capable of forwarding Interests to any entity within.
However, the island does not yet trust SN\textsubscript{2}.
In order to establish this trust, SN\textsubscript{2} begins the second phase, which 
is illustrated in Fig.~\ref{fig:proto_authentication}.
This phase begins with SN\textsubscript{2} sending its \textit{SN Authentication} (SA), a signed
Interest to IM containing the previously exchanged nonces,
R\textsubscript{SN\textsubscript{2}} and R\textsubscript{IM}, as well as
ID\textsubscript{SN\textsubscript{2}}, ID\textsubscript{AN}, and ID\textsubscript{IM}.
This Interest is to be routed using the next-hop information ascertained in the first phase.
The IM, upon receiving the Interest, verifies the signature and content and produces a Data packet
containing the anchor-specific RAK\textsubscript{AN} (shared secrets between IM and ANs are always synchronized).
The key is encrypted under
TEK\textsubscript{SN\textsubscript{2}} and signed under
TAK\textsubscript{SN\textsubscript{2}}.
At this point, SN\textsubscript{2} is authenticated and can move into the third phase
to advertise its path.

\subsection{Path Advertisement}
\label{subsec:publish}

All information necessary for SN\textsubscript{2} to route Interests to other nodes in the island was acquired in the first phase; however, no node is yet able to route interests to SN\textsubscript{2}.
In order for Interests to be delivered to SN\textsubscript{2}, each node on the path between
SN\textsubscript{2} and AN must know the next hop toward SN\textsubscript{2}.
To update this routing state, SN\textsubscript{2} sends a notification called a \textit{SetNext} message
upstream, signed under RAK\textsubscript{AN}.
To keep track of downstream nodes, each SN and AN maintains a Downstream Forwarding Base (DFB), which
maps a node ID to the next-hop MAC address.
The strategy layer of each node uses the DFB and the FIB to make forwarding decisions
regarding Interests with destinations in the same AN's cluster.
To inform the next-hop node of its location, SN\textsubscript{2} creates the SetNext Interest with its
neighbor's prefix (ID\textsubscript{SN\textsubscript{1}}) and the command \texttt{/set-next}, followed
by its own ID (ID\textsubscript{SN\textsubscript{2}}) and the MAC address of SN\textsubscript{1}'s next-hop toward it (in this
case, MAC\textsubscript{SN\textsubscript{2}}).
This Interest is signed with RAK\textsubscript{AN}.
SN\textsubscript{1} receives this interest, updates its DFB, then constructs a similar Interest
informing the next upstream node that it is the next-hop to reach SN\textsubscript{2}.
This process, illustrated in Fig.~\ref{fig:proto_advertisement}, continues until the packet reaches
the AN. 
When the AN receives this Interest, it updates its DFB and sends a \textit{SetPrefix} notification
to the IM to record that it serves as SN\textsubscript{2}'s anchor.
This allows the IM to serve name resolution requests for SN\textsubscript{2}.
The IM responds with a simple ACK message, which should be forwarded hop-by-hop
to satisfy the PIT entries for these Interests, and ultimately notify SN\textsubscript{2}
that it has been successfully onboarded.
Upon receiving the ACK, SN\textsubscript{2} may send a wakeup Interest (Name \texttt{/wakeup}) to notify
nearby nodes that it has been onboarded and can now facilitate their onboarding.
This procedure can help expedite the initial onboarding process for an island.

\subsection{Additional Considerations}
The above protocols accomplish secure onboarding and routing. 
In what follows, we will discuss some additional maintenance procedures in LASeR, such as key refresh, prefix resolution, 
and routing between ANs.

\subsubsection{Key Refresh}

Both the SN's session keys and the AN's RAK may need to be periodically refreshed in order to
maintain the security of the island.
When the SN wants to change keys, it can either restart from the discovery process, or contact
the IM directly to exchange new nonces.
In the latter case, the same authentication procedure applies.
In order to refresh RAK\textsubscript{AN}, the IM should generate the new key and send
\texttt{[\{<RAK\textsubscript{AN}>\}\textsubscript{TEK\textsubscript{i}}]\textsubscript{TAK\textsubscript{i}}}
to each node $i$ in the AN's cluster, as well as the AN itself.

\subsubsection{Prefix Resolution}
\label{subsec:prefix_res}

To enable hierarchical forwarding based on ANs, after committing to a path, an SN assumes a new name-prefix
rooted under its AN.
This prefix is communicated to the IM at the end of the path-advertisement protocol
(Section~\ref{subsec:publish}).
The IM stores a mapping of IDs to prefixes, and can respond to Interests
querying for the prefixes of registered nodes (e.g., respond to an Interest \texttt{/<ID\textsubscript{IM}>/get-prefix/<ID\textsubscript{SN}>)}.
Similarly, the SNs and the ANs need to be able to remap Interests onto the appropriate
prefixes.
Additional arbitrary Name prefixes that a node wishes to serve should also be communicated
to the IM.
The IM also resolves requests for these arbitrary Names into routable Names, constructed from the corresponding IDs and AN prefixes.
The processes for name registration and resolution are quite simple and can be done in various
ways. 
Due to space constraints this discussion is omitted in this paper.

\subsubsection{Routing Within and Between Clusters}
\label{subsec:forwarding}

In LASeR, anchors obtain routes to each other out-of-band, for example using a
link-state algorithm such as NLSR~\citep{hoque2013nlsr}.
By virtue of hierarchical naming and prefix-based forwarding, anchors need not advertise the
constituent nodes in their clusters, only their own prefixes.
We assume that the gateway is also a member of this link-state session, and therefore any
standard node can reach the gateway just as easily as it can reach an anchor.
In order to reach a node within the same cluster, forwarding nodes use their DFBs.
However, each node only has DFB entries for those nodes that rely on it for a
path to the anchor.
Therefore, in our framework nodes route toward the anchor by default if the DFB lookup fails.
As a result, packets between nodes in the same cluster (e.g., machine-to-machine flows) climb
the tree toward the AN until the first common ancestor is reached; the packet is then forwarded
down the tree to the destination.
To reach a node in another cluster, the IM must first be queried to obtain the node's prefix, as
mentioned in Section~\ref{subsec:prefix_res} (Prefix Resolution).
This operation can be made transparent to applications by performing it in the strategy layer, and
the prefix returned by the IM can be cached to reduce latency for future requests.

\subsection{Security Analyses}

The goal of LASeR is to secure routing in a smart city IoT network. 
Therefore, we focus on analyzing concerns related to the authenticity and provenance of obtained routes.
We will also remark on privacy concerns where applicable.
The authentication and key exchange procedures in LASeR are effectively equivalent to those
of EAP-PSK and AKEP2, which have been proven in literature to be secure.
However, the use of a shared RAK for securing routing messages between nodes in a cluster is a potential vulnerability.
Though the RAK is always transmitted in an encrypted form, its compromise (through
brute force or node takeover) would allow an attacker to publish fake routes
on behalf of any node in the compromised cluster.
The result of such an attack would essentially be denial of service of requests destined
to that node (assuming the flow's own data is authenticated). 
This attack can cause blackhole, sinkhole, or wormhole attacks. 
The LASeR protocol could be augmented to report any changes in the topology to the IM, giving
it complete knowledge of the network.
Sophisticated algorithms for detecting blackholes, sinkholes, or wormholes can then be employed
at the IM.
Once detected, the compromised SNs can be revoked and the RAK can be securely refreshed for the remaining 
legitimate SNs. 
In many IoT onboarding protocols, the compromise of an already-trusted node can have major
impacts.
In LASeR, this would result only in the attacker gaining knowledge of the RAK and the node's
own PSK, which it can use to inject fake routes as described above; any application-specific
information obtained would not impact routing security.
If the node is identified, its PSK can be de-authenticated by the IM and the RAK refreshed as
usual.
Any unencrypted data (such as IDs) collected by the compromised node do not
undermine the security of the network; however, they may be used for traffic analysis and
privacy attacks.
Ephemeral IDs and pseudonyms can be used to prevent such information leakage.
Some information about the network topology is leaked by LASeR---a passive attacker could
determine the layout of the network by observing Onboarding Request packets en route to
the IM.
Though we have chosen not to prioritize the protection of this information, the Onboarding
Request and its reply could easily be encrypted under an additional key derived from the 
KDK.
Similarly, the RAK can be augmented with an additional key to enable the encryption of
SetNext and SetPrefix messages.
Node anonymity can also be compromised by observing IDs in transmission; however, because
we allow IDs to be chosen arbitrarily they can be made ephemeral to thwart related attacks.
Channel jamming and other link-layer or physical-layer denial-of-service attacks are beyond the 
scope of our framework.

\section{Simulation Evaluation}
\label{sec:results}

\subsection{Scenario Configuration}

Our initial validation of LASeR was done in ndnSIM~\citep{mastorakis2015ndnsim}, an ns-3
extension implementing NDN.
The implementation of LASeR involves an application-level controller,
a custom forwarding strategy, a modified PIT, and a modified Face which
supports ad-hoc forwarding.
We have also implemented a hop-by-hop fragmentation and reassembly protocol similar
to that in NDNLP~\citep{shi2012ndnlp}.

\begin{figure*}[]
\centering

\subfigure[
\label{fig:convergence_density}
Empirical CDFs of node convergence time.
]{
\includegraphics[width=.3\textwidth,trim={0 0.185cm 0 0.125cm}, clip]{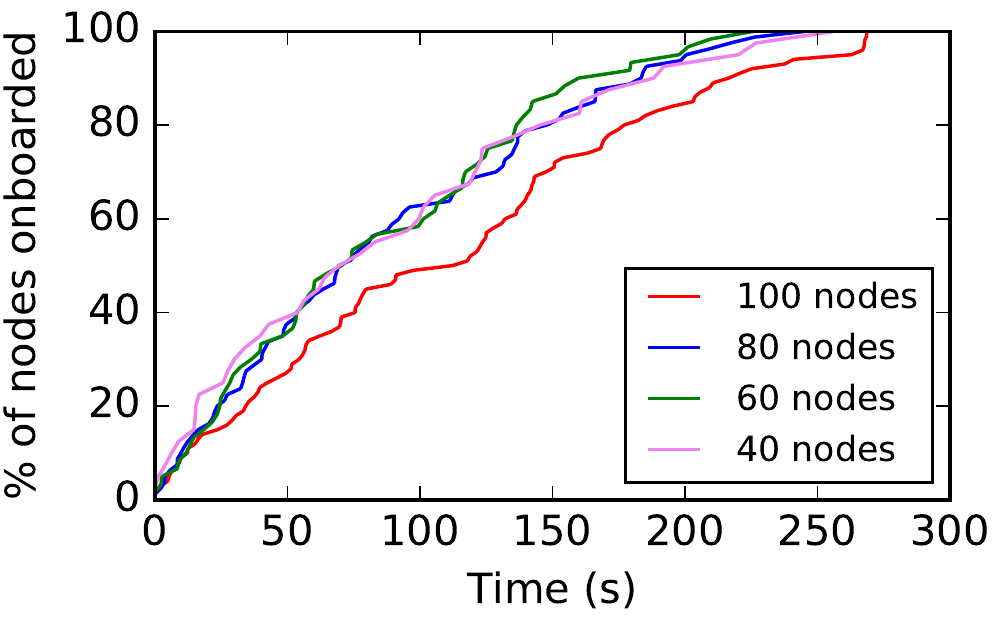}
}
\quad
\subfigure[
\label{fig:transmission_density}
Transmission burdens by subtree size, with standard-error-of-mean.
]{
\includegraphics[width=.3\textwidth,trim={0 0.185cm 0 0.125cm}, clip]{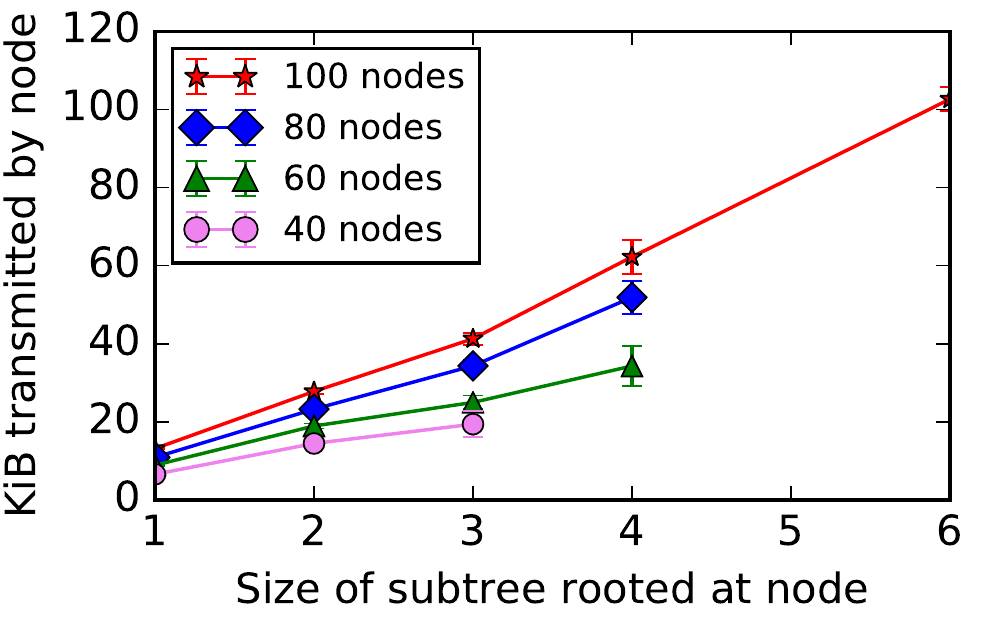}
}
\quad
\subfigure[
\label{fig:subtree_density}
Probability mass function of subtree sizes.
]{
\includegraphics[width=.3\textwidth,trim={0 0.185cm 0 0.125cm}, clip]{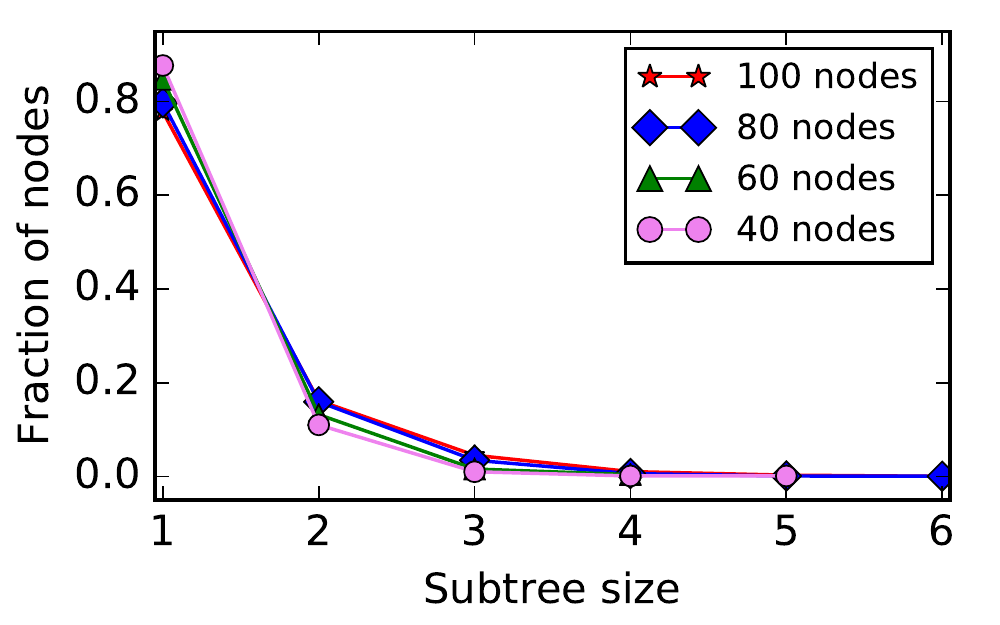}
}

\caption{
Simulation results for the four scenarios with increasing deployment density within a $50\times50$ meter area.
}

\end{figure*}

We used the LrWpanNetDevice to model 802.15.4 radios with slotted CSMA/CA, with
the \textit{Log Distance Propagation Loss} and
\textit{Constant Speed Propagation Delay} models to simulate the radio channel.
%
%
Unfortunately, the practicality of executing large-scale wireless scenarios in ns-3 is limited (as interference calculations become prohibitively expensive),
so we focus on a cluster of SNs around a single AN in each experiment.
The cluster forms the building block of any IoT network, which is essentially composed of
several such clusters.
Onboarding in a single cluster is representative of that in all clusters, as they can happen in parallel.
Therefore, we have not modeled the interconnections between anchors, nor a gateway to a WAN,
and we assume that each anchor is capable of acting on behalf of
the IM.
We do not implement the IM's prefix resolution service for this validation study.
In our evaluations, we explore two settings: increasing density of nodes within a fixed area, and
decreasing density of a fixed number of nodes.
Two sets of scenarios were created for these two settings; they will be detailed in their
respective subsections.
For each scenario, we will explore \textit{four statistics}: time for onboarding convergence,
the transmission burden of each node, the size of each node's subtree, and the hop-count
distances between nodes and their anchors.
Convergence times and hop counts serve as indicators of scalability, while transmission
burdens and subtree sizes correspond to the energy efficiency of the protocol.

\subsection{Increasing Density}

To study the effects of increasing node density, we created scenarios wherein varying
numbers of SNs are placed uniformly at random in a $50\times50$ m$^2$ ($0.0025$ km$^2$), and 
a single AN is placed at the center.
%
%
We evaluated scenarios of $40$ nodes, $60$ nodes, $80$ nodes, and $100$ nodes; this
corresponds to densities ranging from $16,000$ to $40,000$ nodes/km$^2$.
For each scenario, we averaged results over $20$ runs with different pseudo-random
number generator (PRNG) seeds; note that the seed affects both node placement and
network behavior.
To simulate real-world deployments the SNs power-on at random times in the network. 
The time follows an exponential distribution with $\lambda^{-1} = 120$~seconds (2 minutes).  
An SN attempts to join the network after it is powered-on.
%
%

\subsubsection{Convergence Time}

Fig.~\ref{fig:convergence_density} depicts the empirical cumulative distribution functions (eCDFs)
of network convergence times; the X-axis represents time, while
the Y-axis gives the cumulative proportion of nodes that have been onboarded.
As SNs power on randomly, we identified that congestion is not a big challenge to onboarding--new nodes 
get onboarded rapidly.
No clear trend can be seen between the $40$, $60$, and $80$ node scenarios.
However, the $100$-node scenario clearly converges slower, suggesting that as density increases, radio interference has increasingly adverse impact on onboarding.
We believe that with greater densities, onboarding may not
converge.
The worst-case convergence time across all runs was $314.6$ seconds (5 minutes, 14.6 seconds), not
much longer than the average case for $100$ nodes, $271.0$ seconds (4 minutes, 31.0 seconds).
We believe this to be an acceptable convergence delay, as the process only occurs once.
%
%

\subsubsection{Transmission Burden}

The amount of energy consumed by a node is dominated by wireless transmissions.
In LASeR, the transmission burden of an SN grows as it serves increasing number of
other SNs as a forwarder in the onboarding process. 
Therefore, we evaluate the transmission burden observed for SNs of varying subtree sizes.
Application-introduced bandwidth is not considered here; only LASeR traffic is measured.
Anchor nodes are not included in this analysis, as we assume they are not subject
to power constraints.
Fig.~\ref{fig:transmission_density} summarizes our analysis of this transmission burden 
under increasing subtree size and network node density; 
subtree sizes are on the X-axis, and total KiB transmitted is on the Y-axis.
A clear linear trend is visible under increasing subtree size; as expected, transmission
burden is approximately proportional to the number of nodes being served in the subtree.
Additionally, notice that increasing node density results in overall larger transmission
burdens, a pattern resulting from interference-related retransmissions.
In our simulations, a single node must transmit an average of $9.89$ kibibytes (KiB) 
throughout the onboarding process; while this is a reasonable burden, we notice that this burden is  
compounded by increasing the number of downstream SNs.
For this reason, care must be taken to avoid creating large subtrees 
in a real IoT deployment.

\subsubsection{Subtree Size and Distance from Anchor}

%
Fig.~\ref{fig:subtree_density} depicts the empirical probabilities of each observed
subtree size; the X-axis gives subtree sizes, and the Y-axis gives the likelihood that
a node would host a subtree of that size.
On average, 83.6\% of nodes served no children, and thus would experience the minimum transmission burden.
In accordance with intuition, hop-count distance from an anchor was correlated to subtree size;
on average, 79.2\% of nodes were only one hop away from the anchor.
However, we observed that increasing node density increased the average subtree size and path length, even
though the nodes' physical distances from the anchor were unchanged.
This is because increased density increases interference, thus reducing the effective transmission range of nodes and increasing the chance of
SNs to use intermediate forwarders to reach the AN.

\begin{figure*}[]

\centering

\subfigure[
\label{fig:convergence_dist}
Empirical CDFs of node convergence time.
]{
\includegraphics[width=.3\textwidth,trim={0 0.185cm 0 0.125cm}, clip]{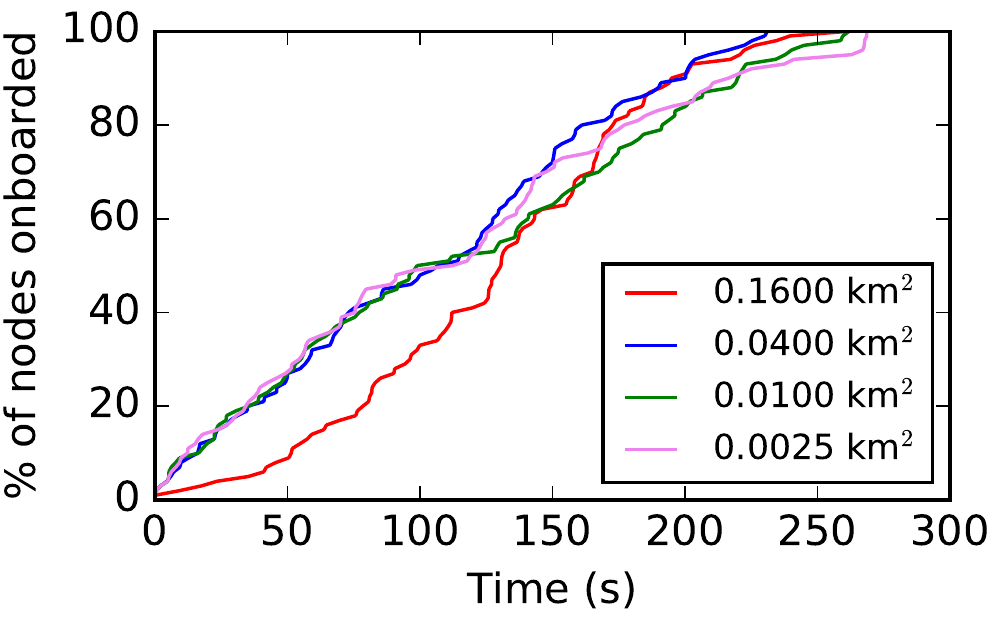}
}
\quad
\subfigure[
\label{fig:transmission_dist}
Transmission burdens by subtree size, with standard-error-of-mean.
]{
\includegraphics[width=.3\textwidth,trim={0 0.185cm 0 0.125cm}, clip]{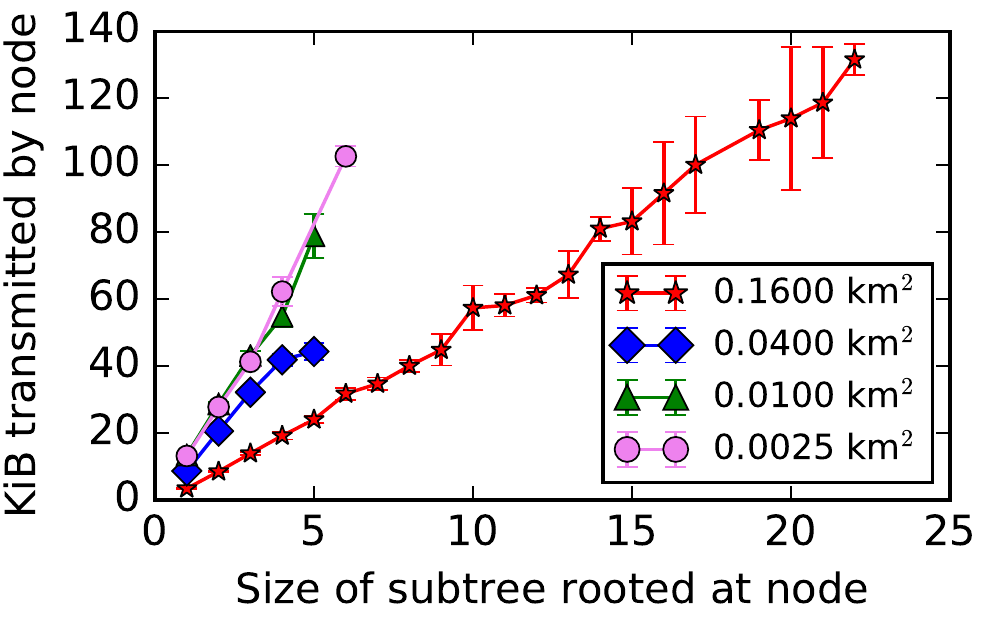}
}
\quad
\subfigure[
\label{fig:subtree_dist}
Probability mass function of subtree sizes.
]{
\includegraphics[width=.3\textwidth,trim={0 0.185cm 0 0.125cm}, clip]{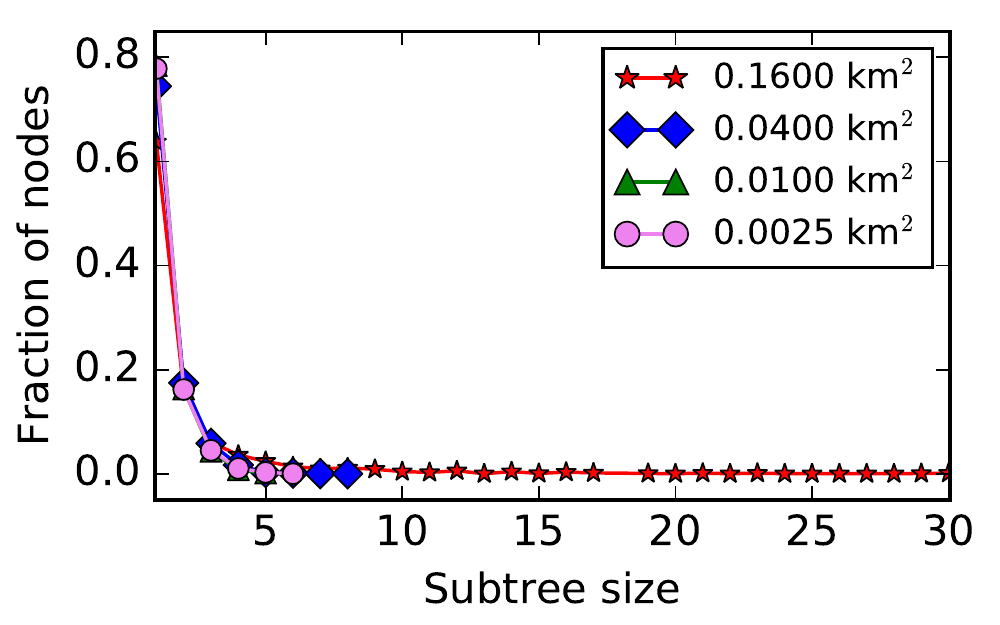}
}

\caption{
Simulation results for the four scenarios of 100 nodes with increasing deployment area.
}

\end{figure*}

\subsection{Increasing Distance}
In the previous subsection, we focused on increasing the number of nodes deployed in a fixed
$50\times50$ m$^2$ area.
We now evaluate a second set of scenarios, wherein the node count is fixed at 100 and
the deployment area is varied.
This increases sparseness, causing creation of longer paths from SNs to the AN and bigger subtrees. 
We chose areas of $50\times50$ m$^2$ (0.0025 km$^2$), $100\times100$ m$^2$ (0.01 km$^2$),
$200\times200$ m$^2$ (0.04 km$^2$), and $400\times400$ m$^2$ (0.16 km$^2$).
%
%
Again, results are averaged over 20 runs.
The time at which nodes come online is exponential with $\lambda^{-1} = 120$~seconds.
With a few exceptions, trends are similar to those observed in the fixed-area scenario set.

\subsubsection{Convergence Time}

The convergence times under the 100-node scenarios are visualized as eCDFs
in Fig.~\ref{fig:convergence_dist}.
We again see that the densest scenario reaches final convergence slowest, however it is
notable that the sparsest scenario ($0.16$ km$^2$) has a slower initial progression.
An inflection point is visible between 100-120 seconds, suggesting that connectivity is
poor prior to a sufficient number of nodes (which will serve as intermediate forwarders) being
onboarded.

\subsubsection{Transmission Burden}

The transmission burdens for nodes with each observed subtree size are visualized in
Fig.~\ref{fig:transmission_dist}.
Again, there is a clear trend of increased burden under higher densities, when considering
similarly-sized subtrees; this is due to interference-related retransmissions.
However, we can see that in the sparsest scenario, some nodes host much larger subtrees
and thus carry a greater burden.
The available energy at these nodes serves as the bottleneck for communication from the downstream nodes.
Thus, care must be taken in node placement: nodes' distances from anchors should be minimized.

\subsubsection{Subtree Size and Distance from Anchor}
The observed probability mass function of subtree sizes is given in Fig.~\ref{fig:subtree_dist}.
Again, a majority of nodes host no children; however, in the sparsest case, a few nodes hosting large
subtrees are observed.
The emergence of large subtrees can be averted by careful node and anchor placement.
The distributions of hop-counts in the three densest scenarios are similar to those observed in the
fixed-area scenario set.
However, in the sparsest scenario a majority of nodes are two hops from an anchor, rather than one hop.
This increases latency; thus, it is best to have short paths from SNs to their corresponding AN.

\section{Conclusions and Future Work}
\label{sec:conclusion}

In this paper, we have proposed LASeR, a secure onboarding and routing framework for
NDN-based IoT networks.
Scalability is achieved through a hierarchical network design, and very little
cryptographic or computational burden.
Evaluation by simulations confirmed that LASeR requires minimal network overhead
and achieves acceptable onboarding convergence times.
The current implementation of LASeR routes based on node IDs, however an extension is
planned to support the advertisement of arbitrary name prefixes.
A mechanism to address node mobility with low overhead is also in development.
After further validating LASeR in ndnSIM, we intend to implement it on real IoT
devices for a live testbed deployment.

\small
\balance
\setlength{\bibsep}{1.86pt}
\bibliographystyle{unsrtnat}
\bibliography{references}

\end{document}